\documentclass[a4paper,10pt]{article}
\usepackage{pos}

\title{Going beyond the Standard paradigm of Cosmology: Torsion, gravitational anomalies and inflation without inflaton fields}

\ShortTitle{Beyond the $\Lambda$CDM: torsion, anomalies and inflation}

\author*[a,b]{Nick E. Mavromatos}

\affiliation[a]{Department of Physics, School of Applied Mathematical and Physical Sciences, National Technical University of Athens, 9 Iroon Polytechniou Str., Zografou Campus GR 157 73, Athens, Greece}

\affiliation[b]{Theoretical Particle Physics and Cosmology Group,
Physics Department,\\
King’s College London,
Strand, London WC2R 2LS}

\emailAdd{nikolaos.mavromatos@cern.ch}

\abstract{I review a string-inspired cosmological model, with gravitational anomalies present at very early epochs, which includes a totally antisymmetric torsion, that in (3+1)-dimensions is equivalent to a pseudoscalar (Kalb-Ramond, string-model independent) axion field. Upon condensation of primordial gravitational waves (GW), that are created at a pre-inflationary era, the model leads to inflation of the so-called running-vacuum-model (RVM) type, which is realised without the need for external inflaton fields, being due to the non-linearities that characterise the gravitational theory.
The model provides an alternative to the $\Lambda$CDM paradigm, and is also capable of inducing matter-antimatter asymmetry in the Universe, during the post-inflationary radiation era, in models with right-handed neutrinos in their matter spectra. The so-induced asymmetry is triggered by a Lorentz-symmetry-violating KR axion background that is generated during the RVM-inflationary period, as a consequence of the GW condensate. The modern era of this cosmology is argued to be characterised by observable (RVM-type) deviations from $\Lambda$CDM, with the potential of alleviating the observed tensions in the cosmological  data.}

\FullConference{%
  7th Symposium on Prospects in the Physics of Discrete Symmetries (DISCRETE 2020-2021)\\
  29th November - 3rd December 2021\\
 Bergen, Norway}

\begin{document}
\maketitle

\section{Introduction and Motivation}

The $\Lambda$CDM paradigm, that is, a cosmological framework based on Einstein's General Relativity (GR) 
with a positive (de-Sitter type) Cosmological Constant $\Lambda > 0$ and a Cold Dark Matter component in its matter sector, is extremely successful in fitting the plethora of the available cosmological data, such as Supernovae type IA, Cosmic Microwave Background (CMB), Baryon Acoustic Oscillations and Weak and Strong lensing data~\cite{Planck}. 
Nonetheless there may be several reasons, motivated by both theory and phenomenology, to go beyond it. 
Theoretically speaking, the theory of quantum gravity is unknown at present. It is possible that a full quantum gravity framework, although encompassing GR at its classical low-energy limit, involves significant corrections to the GR at higher energy scales, such as, for instance, modified gravitational dynamics encoded in higher curvature (and in general higher-gravitational-covariant-derivative) terms. This is the case, for instance, of gravitational theories based on 
string theory~\cite{string2}, which we shall restrict ourselves upon in this review.
From a phenomenological point of view, despite its success in fitting the data, the $\Lambda$CDM paradigm is in tension~\cite{tensions} with cosmological data associated with the local measurement of the current-era Hubble parameter (the so-called $H_0$ tension~\cite{H0,H02,H03}), but also with structure-formation data (the so-called $\sigma_8$ or $S_8$ tensions~\cite{sigma8,sigma83}). Although it is still unclear whether such tensions  point towards fundamental new physics or admit more mundane astrophysical explanations~\cite{mund4}, nonetheless their persisting nature prompted a significant theoretical effort to go beyond the $\Lambda$CDM framework in an attempt to offer potential resolutions~\cite{mena}.

In this talk, which is based on material presented in \cite{Mav1,Mav2}, I will additionally motivate going beyond 
$\Lambda$CDM by the desire: (i)  to answer microscopically the question whether inflation~\cite{inflation} is due to external fundamental scalar inflaton fields, or, like the phenomenologically successful Starobinsky model~\cite{staro},
is a consequence of non-linear gravitational dynamics, which may involve a hidden scalar mode, and (ii) to  
understand in a {\it geometrical} way the observed matter-antimatter asymmetry 
in the Universe~\cite{pdg}, which in the conventional particle physics framework is believed to be due to CP Violation
in the early Universe, along with Baryon-number and C violation, as well as departure from thermal equilibrium (the celebrated Sakharov's conditions~\cite{sakharov}). 

I will adopt a string-inspired gravitational framework, which involves gravitational anomalies in the very early epochs of the Universe, and demonstrate that it entails~\cite{bms1,bms2,ms1,ms2} an early-universe inflationary phase without inflaton fields, of the so-called Running-Vacuum-Model (RVM) type~\cite{rvm1,rvm2,rvmfoss,solaqft,rvmevol1,rvmevol2}, and moreover leads to 
spontaneous Lorentz (and CPT) symmetry violation in the vacuum, through the emergence of Lorentz(LV)- and CPT(CPTV)- Violating  axion backgrounds. The axions are associated with the fundamental massless bosonic string gravitational multiplet, and in particular with the antisymmetric tensor Kalb Ramond (KR) field~\cite{string2}. In the string literature these are known as the string-model-independent (or KR) axion fields~\cite{svrcek}. In the scenario of \cite{bms1,bms2,ms1,ms2}, such LV and CPTV axion backgrounds are created during the RVM inflationary phase, and remain undiluted until the exit phase, and into the radiation era. They are characterised by approximately constant (in cosmic time) rates, and, as such, they can lead to LV and CPTV leptogenesis (lepton asymmetry) in models with right-handed neutrinos in their mass spectra~\cite{decesare,boss}. The lepton asymmetry can then be communicated to the baryon sector (baryogenesis) via baryon-minus-lepton(B-L)-number conserving sphaleron processes~\cite{mssph}, leading to matter-antimatter asymmetry in the Universe. In string theory, the KR axions are associated with a totally antisymmetric torsion in string effective actions~\cite{sloan,metsaev,kaloper}, and in this way the induced matter-antimatter asymmetry acquires a geometric origin. In the post inflationary era, the RVM nature of the cosmology is maintained~\cite{bms1}, and in the modern era it can lead to observable in principle deviations from $\Lambda$CDM~\cite{rvmpheno1,rvmpheno2}, with the potential of alleviating {\it simultaneously} the $H_0$ and $\sigma_8$ tensions~\cite{rvmphenot}. Moreover, as suggested in \cite{bms2}, the KR axions can acquire non-perturbative (instanton-induced) potentials during the post inflationary eras, due to chiral anomalies, which can lead to mass generation, thus making these axion fields potential dark matter candidates.

The structure of this review article is as follows: in the next section \ref{sec:model}, I describe the string-inspired gravitational model and its basic properties. I discuss the r\^ole of the KR axion as torsion, and the microscopic reasons for the appearance of gravitational anomalies in the early epochs of this cosmology. In section \ref{sec:rvminflation}, I describe how an RVM-type inflation arises self-consistently in this model as a result of primordial gravitational-wave (GW) condensation, which in turn leads to a gravitational-anomaly condensation. I demonstrate that the equation of state of the vacuum of this cosmology is that of the RVM model.
In section \ref{sec:postinfl}, I discuss very briefly the post inflationary evolution of this universe, its LV and CPTV matter-antimatter asymmetry, as well as its modern-era  phenomenology, arguing in favour of  the model's potential  to alleviate the observed tensions in the data, and lead to observable in principle deviations from  the $\Lambda$CDM paradigm. Finally, conclusions and outlook are presented in section \ref{sec:concl}.

\section{The String-Inspired Cosmological Model with Gravitational Anomalies and torsion \label{sec:model}} 

The basic assumption of the string-inspired cosmology model of \cite{bms1,bms2,ms1,ms2} is that the early Universe is described by the dynamics of the massless gravitational degrees of freedom only of the underlying string theory model, which in the case of the phenomenologically relevant superstring theory constitute also the ground state. 
The massless fields in the gravitational string multiplet  are the spin-0 dilaton, spin-2 graviton and spin-1 antisymmetric Kalb-Ramond (KR) tensor, and their supersymmetric partners. 
Supergravity is assumed to be broken immediately after the Big Bang, e.g. dynamically via condensation of the supersymmetric partners of gravitons, the gravitinos, as discussed in the model of \cite{ellis}.   In such models there is a first, hill top inflation, not exhibiting necessarily slow roll, which is not observed experimentally. It serves to provide a mechanism for ensuring the washing out of any spatial inhomogeneities and anisotropies before the system settles down in one of the minima of the double-well gravitino condensate potential, which will correspond to the supergravity broken phase. Percolation effects in the early Universe~\cite{lalak,ross} may lead to a lifting of the degeneracy of these two minima of the gravitino-condensate potential, with the 
lower one being the true vacuum, which is reached at the exit from the hill-top inflation~\cite{ms2}.
In this vacuum, the gravitino and its condensate (as well as any other supersymmetric partners of the string gravitational muiltiplet) become massive, with masses close to the Planck mass (which can be arranged by choosing appropriately the supersymmetry-breaking scale $f$ in the model).  
There are instabilities in the hill-top inflationary phase, due to imaginary parts in the respective effective action~\cite{ms2}, which lead to tunnelling to this stable vacuum, in which the supermassive  supersymmetric partners of the bosonic massless gravitational degrees of freedom and the gravitino condensate are integrated out, leaving us with the aforementioned massless dilaton, graviton and KR tensor fields as the propagating dynamical degrees of freedom of this early Universe cosmology. It is this vacuum that will lead to a second, RVM type inflation, with observable consequences, as we shall discuss below~\cite{ms1,ms2}. The dilaton can be self-consistently set~\cite{bms2,ms2} to a constant, and this will be assumed in the remainder of this work.

In the (3+1)-dimensional theory, obtained after string compactification, the axion field is dual to the field strength of the KR antisymmetric tensor spin-1 boson, and in fact it can arise by an implementation in the path integral of the appropriate Bianchi identity for this field strength via a pseudoscalar Langrange multiplier field, which after integrating out the KR spin-1 field, becomes dynamical, and constitutes the KR (or string-model independent) axion field $b(x)$~\cite{kaloper,svrcek}. This theory has in general gravitational anomalies (Lorentz Chern-Simons terms coupled to the axion $b(x)$ field), which arise from the Green-Schwarz (GS) Lorentz Chern-Simons terms in the definition of the modified field strength of the KR field  $B_{\mu\nu}=-B_{\nu\mu}$, $\mu,\nu=0, \dots 3$~\cite{string1,string2} (in form notation for brevity):
$\mathbf{{\mathcal H}} = \mathbf{d} \mathbf{B} + \frac{\alpha^\prime}{8\, \kappa} \, \Omega_{\rm 3L}$, where 
$\Omega_{\rm 3L} = \omega^a_{\,\,c} \wedge \mathbf{d} \omega^c_{\,\,a}
+ \frac{2}{3}  \omega^a_{\,\,c} \wedge  \omega^c_{\,\,d} \wedge \omega^d_{\,\,a}$, with $\omega^a_{\,\,\,b}$  the (torsion-free) spin connection one form (Latin indices are tangent space time indices in a (3+1)-dimensional manifold). 
Above, $\kappa = M_{\rm Pl}^{-1}$ is the gravitational constant in (3+1)-dimensions, with $M_{\rm Pl}=2.4 \times 10^{18}$~GeV the reduced Planck mass, and $\alpha^\prime =M_s^{-2}$, is the Regge slope of the string, with $M_s$ the string mass scale (in units $\hbar=c=1$, we work throughout). In general, $M_s \ne M_{\rm Pl}$, and is a free parameter in string theory, which can be fixed phenomenologically.

We ignored the 
gauge Chern-Simons terms in our scenario, since matter and radiation fields are generated at the end of the second RVM-inflationary period. It can be shown~\cite{sloan,metsaev,kaloper} that the $\mathcal H_{\mu\nu\rho}$, which the effective string-inspired gravitational action depends upon, due to the abelian gauge symmetry $B_{\mu\nu} \rightarrow B_{\mu\nu} + \partial_\mu \theta(x)_\nu - \partial_\nu \theta(x)_\mu$ that characterises the closed string sector~\cite{string1,string2}, can be viewed as a totally antisymmetric torsion, and this interpretation holds up to and including 
$\mathcal O(\alpha^\prime)$ terms in the effective action. For a general discussion and comparison of this model with others in contorted geometries we refer the reader to \cite{Mav2}, and references therein.
The GS Chern-Simons countertems, appearing in the definition of $\mathcal H$, are required in string theory for gauge and gravitational anomaly cancellation in the higher-dimensional theory. In the scenario of \cite{bms1,ms1}, the (3+1)-dimensional gravitational anomalies are not cancelled during the early periods of the Universe, including the RVM inflationary period, but they do cancel after the exit from the RVM inflation, by the gravitational anomalies generated by the  {\it chiral} fermionic matter, which itself emerges at the exit from inflation, as a result of the decay of the RVM vacuum~\cite{rvmfoss,rvmevol2}. The cancellation of the gravitational anomalies during the matter and radiation epochs is required to ensure standard cosmology, given that gravitational anomalies do contribute to the exchange of energy between matter and gravity, in the sense of affecting the conservation of the matter stress  tensor~\cite{jackiw} (in the early universe phase of the cosmology of \cite{bms1,ms1,ms2} the KR axion $b(x)$ belongs to the gravitational multiplet of the string, and hence an exchange of energy between $b(x)$ and the gravitational sector is legitimate (at any rate we note that there is no issue of the validity of general covariance, as discussed in \cite{bms2,ms1})). Chiral anomalies of gauge fields, on the other hand,  survive this cancellation, as they do not contribute to the stress tensor, and hence do not affect energy conservation of matter/radiation. In fact, they may lead to non-perturbative mass generation of the KR axion during the post inflationary eras~\cite{bms2}, implying its potential r\^ole as a dark matter component.

The gravitational action, which is the basis of our string-inspired cosmological model, is then given (to lowest order in an expansion in $\alpha^\prime$ which suffices for our purposes here): 
\begin{align}\label{sea4}
S^{\rm eff}_B &=
\; \int d^{4}x\, \sqrt{-g}\Big[ -\dfrac{1}{2\kappa^{2}}\, R + \frac{1}{2}\, \partial_\mu b \, \partial^\mu b  +  \sqrt{\frac{2}{3}}\,
\frac{\alpha^\prime}{96 \, \kappa} \, b(x) \, R_{\mu\nu\rho\sigma}\, \widetilde R^{\mu\nu\rho\sigma}
   + \dots \Big]
\nonumber \\
& = \; \int d^{4}x\, \sqrt{-g}\Big[ -\dfrac{1}{2\kappa^{2}}\, R + \frac{1}{2}\, \partial_\mu b \, \partial^\mu b  -
 \sqrt{\frac{2}{3}}\,
\frac{\alpha^\prime}{96 \, \kappa} \, {\mathcal K}^\mu (\omega)\, \partial_\mu b(x)   + \dots \Big],
\end{align}
where we used the fact that the Chern-Simons gravitational anomaly terms are total derivatives $R_{\mu\nu\rho\sigma} \, \widetilde R^{\mu\nu\rho\sigma} = \mathcal K^\mu(\omega)_{\,;\,\mu}$, with $;$ denoting the gravitational covariant derivative with respect to the (torsion-free) spin connection $\omega$ (it can be shown that $\mathcal H$-torsion terms can be removed from the anomaly by appropriate counterterms, see discussion and references in~\cite{Mav2}). The tilde denotes the gravitational dual of the Riemann tensor. For notations and conventions the reader is referred to \cite{bms1,ms1}.

We now remark that the Chern-Simons gravity of the form \eqref{sea4} admits Kerr-black hole solutions, with the axion field $b(x)$ being the source of the rotation~\cite{kerrBH,kaloper}, and serving as a provider of axion hair. In the presence of such black holes, the gravitational anomaly terms are therefore non trivial. The same is true when one has GW, which could be the result of either coalescence of such primordial Kerr black holes~\cite{Mav2}, or the non-spherical collapse of unstable domain walls~\cite{ms1} arising from the non-degenerate gravitino-condensate potential due to percolation effects, as mentioned above~\cite{lalak,ross}. When such GW are produced in abundant quantities, they may condense, which 
can then lead to inflation of an RVM type, without inflaton fields, as we now proceed to discuss~\cite{ms1,ms2}.

\section{Running-Vacuum-Model Inflation without Inflaton fields \label{sec:rvminflation}}

In the presence of GW perturbations, the gravitational anomaly term is non  trivial~\cite{stephon}. In \cite{bms1,ms1}, we first assumed an inflationary phase, with an approximately constant Hubble parameter,  
and calculated the pertinent anomaly condensate due to GW condensation, following the calculation of \cite{stephon}. 
Then we demonstrated that the condensation of the gravitational anomaly implies a RVM type energy density, charactereised by an RVM equation of  state~\cite{rvm1,rvm2,rvmfoss,solaqft}. It is known that in the RVM framework inflation arises dynamically at the early stages of the Universe due to non-linearities in the evolution equation of the RVM universe~\cite{rvmevol1,rvmevol2}. In this way, we then arrive at the self consistency of the main point of our analysis, namely that the GW condensation leads to an RVM inflation without external inflaton fields. 

Let us review briefly the basic steps of our approach~\cite{bms1,ms1,ms2}. 
Assuming a GW perturbation with metric, say: $ds^2 = dt^2 - a^2(t) \Big[(1 - h_+(t,z))\, dx^2 + (1 + h_+(t,z))\, dy^2 + 2h_\times (t,z)\, dx\, dy + dz^2 \Big]$,  with the scale factor of the Universe being given by $a(t) \sim e^{H_I\, t},$ where $H_I$ is approximately constant, corresponding to the inflation scale, we evaluate the gravitational anomaly condensate, following \cite{stephon}, by integrating over graviton modes with momenta up to an UltraViolet (UV) cut-off scale $\mu$:
\begin{align}\label{rrt2}
  \langle R_{\mu\nu\rho\sigma}\, \widetilde R^{\mu\nu\rho\sigma} \rangle  &\simeq  \frac{d}{dt} \mathcal K^0(t) = 
  \frac{16}{a^4} \, \kappa^2\int^\mu \frac{d^3k}{(2\pi)^3} \, \frac{H_I^2}{2\, k^3} \, k^4 \, \Theta  = \frac{1}{\pi^2} \Big(\frac{H_I}{M_{\rm Pl}}\Big)^2 \, \mu^4\, \Theta   \nonumber \\
 &= \frac{2}{3\pi^2} \frac{1}{96 \times 12} \,  \Big(\frac{H_I}{M_{\rm Pl}}\Big)^3 \, \Big(\frac{\mu}{M_{\rm Pl}}\Big)^4 \,  M_{\rm Pl}\, \times \, \, {\mathcal K}^0 (t)\,,
\end{align}
to leading order in the slow-roll parameter $ \Theta = \sqrt{\frac{2}{3}}\, \frac{\alpha^\prime \, \kappa}{12} \, H_I \,  {\dot {\overline b}} \, \ll \, 1$,
with the overdot denoting derivative with respect to the cosmic time $t$ in the Robertson-Walker frame. We also assumed homogeneity and isotropy in the Universe, which is guaranteed in our model by the existence of a first Hill-top inflation~\cite{ellis}, arising from dynamical supergravity breaking, as discussed in the previous section.  This is manifested in \eqref{rrt2} by the dominance of the temporal component of the anomaly $\mathcal K^0(t)$. For details we refer the reader to \cite{bms1,bms2,ms2}. The equation \eqref{rrt2} leads to an evolution equation for $\mathcal K^0$, which, upon an appropriate choice of parameters in the model, namely $\frac{\mu}{M_s} \simeq 15 \, \Big(\frac{M_{\rm Pl}}{H_I}\Big)^{1/2} \simeq 10^3$ (using the Planck data for the inflationary scale $H_I \simeq 10^{-5}\, M_{\rm Pl}$~\cite{Planck}), leads to an approximately constant anomaly $\mathcal K^0(t) \simeq {\rm constant}$, for the entire duration of inflation. From the equations of motion of the $b(t)$ field then, stemming from \eqref{sea4}, we obtain (spontaneous) Lorentz and CPT violation through the axion background solution:
\begin{align}\label{krbeom2}
\dot{\overline{b}}  =  \sqrt{\frac{2}{3}}\, \frac{\alpha^\prime}{96 \, \kappa} \, {\mathcal K}^{0} \simeq {\rm constant} 
= \sqrt{2\epsilon} \, H_I \,  M_{\rm Pl} \,\, \Rightarrow \,\,  \overline b(t) = \overline b(t_0) + \sqrt{2\epsilon} \, H_I  \, M_{\rm Pl} \, (t - t_0),
\end{align}
where the right-hand-side of the first equation in \eqref{krbeom2} is a convenient parameterisation, which agrees with the slow-roll CMB data~\cite{Planck} for $\epsilon = {\mathcal O}(10^{-2})$, since it is the axion field $b$ that drives the slow-roll in this case~\cite{bms1,ms1}. The quantity $\overline b(t_0)$ is a boundary condition for the KR axion field at the onset of the second (RVM) inflation, at $t=t_0 > 0$ (assuming the Big Bang occurring at $t=0$). 
The requirement of the satisfaction of the transplanckian conjecture, that is, no modes with momenta higher than the Planck scale appear in the spectrum, including gravitons, and hence the UV cutoff $\mu$ in \eqref{rrt2} is at most of order of the Planck scale, in conjunction with the slow-roll condition for the $b(x)$ axion, imply the allowed range of the string mass scale~\cite{Mav2,ms1}: $2.6 \times 10^{-5}\, M_{\rm Pl}  \ll M_s \le 10^{-3}\, M_{\rm Pl}$. On average, one may therefore take $M_s = \mathcal O(10^{-3}) \, M_{\rm Pl}$, implying $\mu =\mathcal O(M_{\rm Pl})$, which is natural.
As discussed in \cite{bms1,ms1}, on assuming in \eqref{krbeom2} that $ \frac{|\overline b(t_0)|}{M_{\rm Pl}}  \gg \, \sqrt{2\, \epsilon} \,  \mathcal N = \mathcal O(10), \, \overline b(0) < 0$, where $\mathcal N $ are the e-foldings, which should be at least of order $\simeq 55-60$ for phenomenologically consistent inflation~\cite{inflation,Planck}, one ensures an approximately  constant  condensate $\mathcal F \equiv \langle \overline b(t_0) \, R_{\mu\nu\rho\sigma}\, \widetilde R^{\mu\nu\rho\sigma} \rangle \simeq 5.86 \times 10^{-5}  \, \Big(\frac{\mu}{M_s}\Big)^4 \, \sqrt{2\, \epsilon} \, \Big[\frac{\overline b(t_0)}{M_{\rm Pl}}\Big] \, H^4$, which leads to a positive cosmological-constant-like (de Sitter) term in the effective action \eqref{sea4}, and hence inflation. One can expand the effective action \eqref{sea4} about such condensates.
It can then be shown~\cite{ms1,ms2}, upon taking into account the aforesaid condition $\mu = \mathcal O(10^{3}) \, M_s \simeq M_{\rm Pl}$, that the total energy density of this fluid, comprising of contributions from KR-axion-$b$-parts, gravitational Chern-Simons fluctuations and the condensate $\mathcal F$, assumes an RVM form:
\begin{align}\label{toten}
0 <  \rho_{\rm total}  \simeq  3\kappa^{-4} \, \Big[ -1.65 \times 10^{-3} \Big(\kappa\, H \Big)^2
+ \frac{\sqrt{2}}{3} \, |\overline b(t_0)| \, \kappa \, \times {5.86\, \times} \, 10^6 \, \left(\kappa\, H \right)^4 \Big]
\end{align}
under the aforementioned condition for the magnitude of $|\overline b(t_0)|$. The overall positive nature of the total energy density is due to the dominance of the condensate $H^4$ term. On the other hand, the Chern-Simons terms contribute negatively to the energy density, and are responsible for the negative value of the coefficient of the $H^2$ terms.\footnote{Par contrast, in conventional quantum-field-theory computations in curved spacetime, the dominant at late eras $H^2$ terms in the vacuum energy density, with a positive coefficient, can be derived in an effective field theory approach with conventional matter fields~\cite{solaqft}. This also characterises the post inflationary cosmology of the stringy-RVM model of 
\cite{bms1,ms1,ms2}, where the gravitational anomalies are cancelled and ordinary matter and radiation remain ({\it cf.} section \ref{sec:postinfl}).}   In \eqref{toten}, $H \simeq H_I$ is the approximately constant Hubble parameter during inflation. 
The reader should notice the CP-Violating nature of the Chern-Simons gravitational anomaly, as well as that of the GW perturbations, whose condensate leads to inflation. In this sense we see a rather important, but different from
Sakharov's conditions~\cite{sakharov}, r\^ole of CP violation in this early string-inspired Universe.

The form of \eqref{toten} is that of an RVM total energy density, except for the negative value of the coefficient of the $H^2$ terms. It can also be shown~\cite{ms1,ms2}, using properties of the Chern-Simons gravity~\cite{jackiw}, that the total equation of state of this fluid, including contributions from the KR axion ($b$), gravitational Chern-Simons terms (gCS) and the condensate $\mathcal F$, is that of the RVM~\cite{rvm1,rvm2,rvmfoss,solaqft}:
\begin{align}\label{rvmeos}
p_{\rm total} = - \rho_{\rm total}, \quad {\rm where} \quad p_{\rm total} = p^b + p^{\rm gCS} + p^{\mathcal F}, \quad 
\rho_{\rm total} = \rho^b + \rho^{\rm gCS} + \rho^{\mathcal F}, 
\end{align}
with the individual components also satisfying 
$p^b = + \rho^b, \, \, p^{\rm gCS} = \frac{1}{3} \, \rho^{\rm gCS} \,\, \Rightarrow \,\,
p^b + p^{\rm gCS} = \rho^b + \frac{1}{3} \, \rho^{\rm gCS} = -\frac{1}{3}\, \rho^{\rm gCS} = - (\rho^b + \rho^{\rm gCS})>0$, and  $p^{\mathcal F} = - \rho^{\mathcal F} \,< \,0$.
We observe that, were it not for the dominance of the condensate $\mathcal F$, one would face a phantom-matter situation with negative energy ({\it i.e.} violating the weak energy conditions)~\cite{ms2}. The above results hold even for slowly varying Hubble parameter $H(t)$ with the cosmic time, as required by RVM.

Reversing the logic then we may invoke the RVM evolution equation, as described in \cite{rvmevol1,rvmevol2}, appropriately adapted to our problem~\cite{Mav2}, in order to infer the existence of an early de Sitter phase, leading to an approximately constant $H \simeq H_I$, driven by the $H^4$ term in \eqref{toten}, which dominates the early eras. This RVM inflation does not require external inflaton fields, but is a direct  consequence of the non-linearities of the RVM evolution. This provides a non-trivial self consistency check of our analysis. There is a hidden scalar mode, of course, in the condensate $\mathcal F$. The slow-roll parameter of this RVM inflation is provided by the LV and CPTV background solution for the KR axion field \eqref{krbeom2}. This background remains {\it undiluted} at the exit from inflation, and well onto the radiation era, where it can lead to leptogenesis in models involving right-handed neutrinos~\cite{decesare,boss,ms1}, as we discuss in the next section, where we describe briefly the post inflationary era of the model.

\section{Post Inflationary era, matter antimatter asymmetry and late-epochs deviations from $\Lambda$CDM \label{sec:postinfl}}

At the exit from the RVM inflation, chiral fermionic matter is generated, along with other gauge fields. The chiral fermions are characterised by their own gravitational and chiral anomalies, in the sense of the one-loop exact relation
\begin{eqnarray}
   \label{anom}
J^{5\mu}_{\,\,\,\,\,\,\,\,;\mu}  \!= \frac{1}{\sqrt{-g}} \, \partial_\mu \Big(\sqrt{-g} \, J^{5\, \mu}\Big) =  \frac{1}{\sqrt{-g}}\,
\! \partial_\mu \Big(\sqrt{-g} \, \frac{\mathcal N_{\rm ch}}{192 \pi^2} \, \mathcal K^\mu \Big) - 
\frac{e^2\, \mathcal N_{\rm ch}}{32\pi^2} {\mathbf F}^a_{\mu\nu}\,  \widetilde{\mathbf F}^{a\,\mu\nu}, 
\end{eqnarray}  
where $J^{5\mu}$ is the axial current of the chiral fermion species, including right-handed neutrinos in models 
which involve them in their spectra, and $\mathcal N_{\rm ch}$ denotes the chiral degrees of freedom in this string-inspired model, and $\mathbf F$ are the field strengths of, in general, non-Abelian gauge fields (the tilde denotes their dual). The last term on the right-hand side of \eqref{anom} is the chiral anomaly, which, in contrast to the gravitational anomaly term, is topological, not contributing to the stress tensor, and hence not affecting matter-energy-momentum conservation. 
In \cite{bms1,ms1} it was assumed that by adding appropriate GS counterterms to the effective (3+1)-dimensional gravitational action, one can cancel the primordial gravitational anomalies, existing in \eqref{sea4} due to the GS counterterms, by those generated by the chiral fermions. This would be ensured by appropriately fixing $\mathcal N_{\rm ch}$ for a given coefficient of the 
gravitational Chern-Simons term in \eqref{sea4} (equivalently, for a given $\mathcal N_{\rm ch}$, depending on the model, one fixes accordingly the coefficient of the Chern-Simons anomaly in \eqref{sea4}~\cite{Mav1,Mav2}).   Such a cancellation ensures that the conventional Friedmann-Lemaitre-Robertson-Walker (FLRW) cosmology, where matter-radiation energy is conserved,  holds for the post-RVM-inflationary epochs of the model. 

Chiral anomalies, though, remain, and under certain circumstances may lead to a non-perturbative (instanton type) generation of a periodic potential, and thus a mass, for the KR axion field, during, say, some epoch in the post inflationary stringy-RVM universe, e.g. the QCD epoch~\cite{bms2}, in which case the gauge field strength $\mathbf F$ in \eqref{anom} could correspond to the gluon field strength. For the range of the string scale found in our problem, one then obtains phenomenologically acceptable masses for the KR field, which could resemble the QCD axion, thus playing the r\^ole of an acceptable candidate for dark matter. In view of the connection of the $b(x)$ field to torsion, therefore, one may have a geometric origin of dark matter~\cite{Mav1}.  Other scenarios, however, exist, in which the KR axion develops complicated potentials that mix it with the other axions arising in string theory due to compactification~\cite{svrcek}, which can lead to an interesting and rich phenomenology~\cite{bms1}, to be explored further in future works. 

The cancellation of gravitational anomalies implies, on account of the equations of motion following from the 
effective action with chiral matter and radiation at the exit from RVM inflation, that the background axion would now satisfy 
$\frac{1}{\sqrt{-g}}\, \partial_\mu \Big(\sqrt{-g}\, \partial^\mu \overline b \Big) = - {\mathcal O}\Big(\frac{\alpha_{\rm EM}}{2\pi} \,  {F}^{\mu\nu}\,  \widetilde{F}_{\mu\nu}\, , \, \frac{\alpha_s}{8\pi}\, G_{\mu\nu}^a \, \widetilde G^{a\mu\nu} \Big)$,
where the term on the right-hand side is due to the chiral anomaly contributions ({\it cf.} \eqref{anom}), with $F_{\mu\nu}$ the electromagnetic Maxwell tensor, and $G_{\mu\nu}^a$, $a=1, \dots 8$ the SU$_{\rm c}$(3) gluon tensor. In the scenario of \cite{bms2,ms1} the chiral anomalies do not appear during the very early eras after RVM inflation, in which case the above equation for $\overline b(t)$ would have a vanishing right-hand side for very early radiation era, and thus a solution $\dot{\overline b}(t) \sim T^{-3}$, where $T$ is the cosmic temperature, with which the radiation-era scale factor of the universe scales as $a(t) \sim T^{-1}~$. 

Such backgrounds $B_\mu \propto \dot{\overline b} \,\delta_{\mu0}$ couple to the right-handed neutrinos, if the latter exist in the matter spectrum of the model, as assumed in \cite{ms1}, and this leads to LV and CP-Violating (CPV) and  CPTV leptogenesis, according to the mechanism described in \cite{decesare,boss}, due to the axial-current-KR-axion-background coupling $J^{5\mu} B_\mu$ in the respective effective Lagrangian. The lepton asymmetry occurs in this model due to the different decay rates of (Majorana) right-handed neutrinos to standard model leptons and antileptons, as a result of standard Yukawa-type portal interactions that connect the right-handed neutrino sector with the standard model sector. In the presence of the LV and CPTV background, contrary to the standard CPV but CPT conserving, Lorentz invariant leptogenesis case, the rates are unequal already at tree level, and even for a single species  of right-handed neutrinos. In contrast, in the conventional case, to obtain the necessary CPV required for the lepton asymmetry, one needs more than one species of right-handed neutrinos and one-loop diagrams.
For the short period of leptogenesis a background $B_0 \propto T^{-3}$ is considered as approximately constant, 
and one can then apply semi-analytic techniques to estimate the lepton asymmetry~\cite{boss}, which for weak backgrounds $|B_0|/m_N \ll 1$, where $m_N$ is the mass of the right-handed neutrino, turns out to be proportional to the first power of $|B_0|/m_N$~\cite{decesare}. Such asymmetries are then communicated to the Baryon sector 
via B-L conserving sphaleron processes.\footnote{We note at this stage that, since the axion background $B_\mu$, being related to torsion, couples  {\it universally} to all species of chiral fermions, including quarks, through its interaction with the axial current, one could also think of scenarios where direct LV and CPTV baryogenesis can occur. In our stringy RVM~\cite{bms1,ms1}, though, we adopted the baryogenesis through leptogenesis scenario, as it suffices to provide simple and phenoenologically realistic ways of generating the matter antimatter asymmetry in the Universe.} In the model of \cite{boss}, which was adopted in \cite{bms1,ms1}, 
leptogenesis can occur at high temperatures, with freezeout $T_D \simeq m_N \sim 100$~TeV.

It can be shown~\cite{bms1} that during the post inflationary period, a conventional RVM form, with a positive coefficient of the $H^2$ term, characterises the corresponding vacuum energy density (this, {\it e.g.}, can be the result of background cosmic electromagnetic fields). The resulting vacuum energy density at late eras, has the form $\rho_0^{\rm RVM} \sim 3\kappa^{-2}(c_0 + \nu H_0^2 )$, with $c_0 > 0$ the cosmological constant, and $\nu > 0$, where the subscript $``0''$ denotes present-day quantities, in agreement with the quantum-field theoretic derivations of the standard RVM~\cite{solaqft}. Such  vacuum energies lead to observable in principle deviations from $\Lambda$CDM~\cite{rvmpheno1,rvmpheno2}, with a fit to data constraining the (positive) coefficient $0 < \nu = \mathcal O(10^{-3})$. Moreover, they have the potential to alleviate the observed tensions in the data~\cite{rvmphenot}. In fact, as argued in \cite{ms1,Mav1,Mav2}, in the stringy RVM context, quantum graviton fluctuations lead to logarithmically-dependent-on-$H$ corrections to the conventional RVM, so that 
the vacuum energy density assumes the form 
$\rho_0^{\rm stringy RVM} (H) = 3\kappa^{-2} \Big[c_0 + \Big(\nu + d_1\,{\rm ln}(M_{\rm Pl}^{-2} H^2) \Big) \, H^2 + \dots \Big]$, which resembles somehow the type II RVM model, that is capable of resolving simultaneously the $H_0$ and $\sigma_8$ tension~\cite{rvmphenot}. However, in contrast to the type II RVM, in which the effective gravitational constant $\kappa^2_{\rm eff} = \kappa^2/\varphi(t)$ scales mildly with the cosmic time $t$, our stringy-RVM is characterised by a constant $\kappa$ and a constant term $c_0$, {\it i.e.} they do not vary with the cosmic time.

\vspace{-0.39cm}

\section{Conclusions and Outlook \label{sec:concl}}

\begin{figure}[h]
 \centering
  \includegraphics[clip,width=0.7\textwidth,height=0.3\textheight]{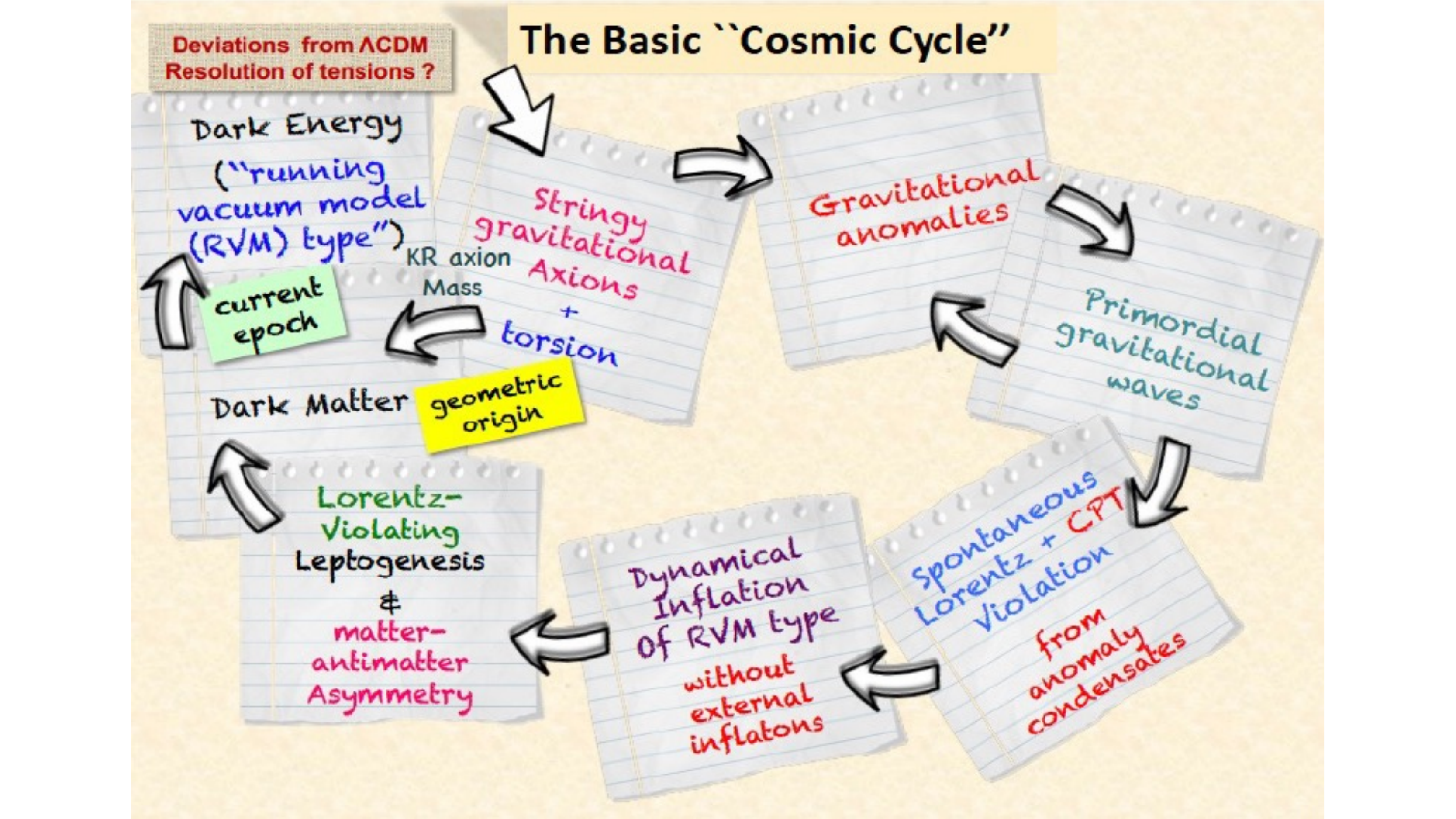} 
\caption{\it Basic cosmic cycle that characterises the string-inspired version of the running-vacuum model, proposed in \cite{bms1,ms1,ms2}. The model includes gravitational anomalies, and totally antisymmetric torsion, which in (3+1)-dimensions is dual to a pseudoscalar (axion-like) field $b(x)$. Upon acquiring a non-perturbative (instanton-induced) mass during the post-RVM-inflationary era, the latter can play the r\^ole of Dark Matter.}\label{fig:cycle}
\end{figure}

Instead of conclusions  we would like to give the basic cosmic cycle characterising the string version of the RVM proposed in \cite{bms1,bms2,ms1,ms2} and reviewed here. This cycle is depicted in fig.~\ref{fig:cycle}. 

As an outlook, we mention that there are several avenues for research that one can pursue in connection with this approach. One, is a further understanding of the current era of the model, and its quantum-gravity-induced logarithmic dependences on the Hubble parameter, as far as their phenomenology is concerned, and their potential to alleviate the tensions. A second, rather ambitious task, is the early universe phenomenology of the Lorentz and CPT Violating axion backgrounds, which may leave detectable imprints in the CMB spectra. A more complicated issue is the phenomenological testing of the Chern-Simons-induced negative coefficients of the $H^2$ terms in the RVM vacuum energy density during the RVM inflationary phase. Since at the exit from this phase there is a phase transition involved, this task is a very difficult one, and at present it is not known to the author how to develop sufficiently accurate techniques to 
address it. Another task is the (rich) phenomenology of other axion fields that arise from string compactification, which mix with the KR axion. As speculated in \cite{Mav2}, in multiaxion scenarios, some of the stringy axions can develop periodic, shift-symmetry breaking, potential structures during the RVM inflation, and can lead to parametric resonant phenomena, that can enhance dramatically the fluctuations of the $b(x)$ axion field, which is assumed to respect the shift symmetry during the RVM inflation, thus producing large curvature fluctuations. The density of primordial black holes, then, that are produced during the RVM inflationary phase, may be enhanced significantly, influencing the profile of the GW, and leading in turn to their dominance over the GW produced during radiation. This could lead to observable, in principle, effects in interferometers.

\section*{Acknowledgements} 

I would like to thank Prof. Per Osland and the local organising committee of the Discrete 20-21 conference for their kind invitation to speak in the plenary session, and for organising such a thought stimulating conference. 
This work is supported in part by the UK Science and Technology Facilities Council (STFC)
under research grant ST/T000759/1. The author acknowledges participation in the COST Association Action CA18108 {\it Quantum Gravity Phenomenology in the Multimessenger Approach (QG-MM)}.

\bibliographystyle{JHEP}
\small
\bibliography{Mavromatos_DISCRETE_20_21}

\providecommand{\href}[2]{#2}\begingroup\raggedright\begin{thebibliography}{10}

\bibitem{Planck}
{\scshape Planck} collaboration, \emph{{Planck 2018 results. VI. Cosmological
  parameters}},
  \href{https://doi.org/10.1051/0004-6361/201833910}{\emph{Astron. Astrophys.}
  {\bfseries 641} (2020) A6}
  [\href{https://arxiv.org/abs/1807.06209}{{\ttfamily 1807.06209}}].

\bibitem{string2}
M.B.~Green, J.H.~Schwarz and E.~Witten, \emph{{Superstring Theory Vol. 2}:
  {25th Anniversary Edition}}, Cambridge Monographs on Mathematical Physics,
  Cambridge University Press (11, 2012),
  \href{https://doi.org/10.1017/CBO9781139248570}{10.1017/CBO9781139248570}.

\bibitem{tensions}
L.~Verde, T.~Treu and A.G.~Riess, \emph{{Tensions between the Early and the
  Late Universe}},
  \href{https://doi.org/10.1038/s41550-019-0902-0}{\emph{Nature Astron.}
  {\bfseries 3} (2019) 891} [\href{https://arxiv.org/abs/1907.10625}{{\ttfamily
  1907.10625}}].

\bibitem{H0}
A.G.~Riess et~al., \emph{{A 2.4\% Determination of the Local Value of the
  Hubble Constant}},
  \href{https://doi.org/10.3847/0004-637X/826/1/56}{\emph{Astrophys. J.}
  {\bfseries 826} (2016) 56}
  [\href{https://arxiv.org/abs/1604.01424}{{\ttfamily 1604.01424}}].

\bibitem{H02}
A.G.~Riess et~al., \emph{{Milky Way Cepheid Standards for Measuring Cosmic
  Distances and Application to Gaia DR2: Implications for the Hubble
  Constant}}, \href{https://doi.org/10.3847/1538-4357/aac82e}{\emph{Astrophys.
  J.} {\bfseries 861} (2018) 126}
  [\href{https://arxiv.org/abs/1804.10655}{{\ttfamily 1804.10655}}].

\bibitem{H03}
V.~Bonvin et~al., \emph{{H0LiCOW \textendash{} V. New COSMOGRAIL time delays of
  HE 0435\ensuremath{-}1223: $H_0$ to 3.8 per cent precision from strong
  lensing in a flat \ensuremath{\Lambda}CDM model}},
  \href{https://doi.org/10.1093/mnras/stw3006}{\emph{Mon. Not. Roy. Astron.
  Soc.} {\bfseries 465} (2017) 4914}
  [\href{https://arxiv.org/abs/1607.01790}{{\ttfamily 1607.01790}}].

\bibitem{sigma8}
E.~Macaulay, I.K.~Wehus and H.K.~Eriksen, \emph{{Lower Growth Rate from Recent
  Redshift Space Distortion Measurements than Expected from Planck}},
  \href{https://doi.org/10.1103/PhysRevLett.111.161301}{\emph{Phys. Rev. Lett.}
  {\bfseries 111} (2013) 161301}
  [\href{https://arxiv.org/abs/1303.6583}{{\ttfamily 1303.6583}}].

\bibitem{sigma83}
E.~Di~Valentino et~al., \emph{{Cosmology intertwined III: $f\sigma_8$ and
  $S_8$}},
  \href{https://doi.org/10.1016/j.astropartphys.2021.102604}{\emph{Astropart.
  Phys.} {\bfseries 131} (2021) 102604}
  [\href{https://arxiv.org/abs/2008.11285}{{\ttfamily 2008.11285}}].

\bibitem{mund4}
W.L.~Freedman, \emph{{Cosmology at a Crossroads}},
  \href{https://doi.org/10.1038/s41550-017-0121}{\emph{Nature Astron.}
  {\bfseries 1} (2017) 0121}
  [\href{https://arxiv.org/abs/1706.02739}{{\ttfamily 1706.02739}}].

\bibitem{mena}
E.~Di~Valentino, O.~Mena, S.~Pan, L.~Visinelli, W.~Yang, A.~Melchiorri et~al.,
  \emph{{In the realm of the Hubble tension\textemdash{}a review of
  solutions}}, \href{https://doi.org/10.1088/1361-6382/ac086d}{\emph{Class.
  Quant. Grav.} {\bfseries 38} (2021) 153001}
  [\href{https://arxiv.org/abs/2103.01183}{{\ttfamily 2103.01183}}].

\bibitem{Mav1}
N.E.~Mavromatos, \emph{{Geometrical origins of the Universe dark sector:
  string-inspired torsion and anomalies as seeds for inflation and dark matter,
  Phil. Trans A (Roy. Soc. UK), in press}},
  \href{https://arxiv.org/abs/2108.02152}{{\ttfamily 2108.02152}}.

\bibitem{Mav2}
N.E.~Mavromatos, \emph{{Torsion in string-inspired cosmologies and the universe
  dark sector}}, \href{https://doi.org/10.3390/universe7120480}{\emph{Universe}
  {\bfseries 7} (2021) 12} [\href{https://arxiv.org/abs/2111.05675}{{\ttfamily
  2111.05675}}].

\bibitem{inflation}
J.~Martin, C.~Ringeval and V.~Vennin, \emph{{Encyclop\ae{}dia Inflationaris}},
  \href{https://doi.org/10.1016/j.dark.2014.01.003}{\emph{Phys. Dark Univ.}
  {\bfseries 5-6} (2014) 75} [\href{https://arxiv.org/abs/1303.3787}{{\ttfamily
  1303.3787}}].

\bibitem{staro}
A.A.~Starobinsky, \emph{{A New Type of Isotropic Cosmological Models Without
  Singularity}},
  \href{https://doi.org/10.1016/0370-2693(80)90670-X}{\emph{Phys. Lett. B}
  {\bfseries 91} (1980) 99}.

\bibitem{pdg}
{\scshape Particle Data Group} collaboration, \emph{{Review of particle
  physics. Particle Data Group}},
  \href{https://doi.org/10.1016/j.physletb.2004.06.001}{\emph{Phys. Lett. B}
  {\bfseries 592} (2004) 1}.

\bibitem{sakharov}
A.D.~Sakharov, \emph{{Violation of CP Invariance, C asymmetry, and baryon
  asymmetry of the universe}},
  \href{https://doi.org/10.1070/PU1991v034n05ABEH002497}{\emph{Pisma Zh. Eksp.
  Teor. Fiz.} {\bfseries 5} (1967) 32}.

\bibitem{bms1}
S.~Basilakos, N.E.~Mavromatos and J.~Sol\`a~Peracaula, \emph{{Gravitational and
  Chiral Anomalies in the Running Vacuum Universe and Matter-Antimatter
  Asymmetry}}, \href{https://doi.org/10.1103/PhysRevD.101.045001}{\emph{Phys.
  Rev. D} {\bfseries 101} (2020) 045001}
  [\href{https://arxiv.org/abs/1907.04890}{{\ttfamily 1907.04890}}].

\bibitem{bms2}
S.~Basilakos, N.E.~Mavromatos and J.~Sol\`a~Peracaula, \emph{{Quantum Anomalies
  in String-Inspired Running Vacuum Universe: Inflation and Axion Dark
  Matter}}, \href{https://doi.org/10.1016/j.physletb.2020.135342}{\emph{Phys.
  Lett. B} {\bfseries 803} (2020) 135342}
  [\href{https://arxiv.org/abs/2001.03465}{{\ttfamily 2001.03465}}].

\bibitem{ms1}
N.E.~Mavromatos and J.~Sol\`a~Peracaula, \emph{{Stringy-running-vacuum-model
  inflation: from primordial gravitational waves and stiff axion matter to
  dynamical dark energy}},
  \href{https://doi.org/10.1140/epjs/s11734-021-00197-8}{\emph{Eur. Phys. J.
  ST} {\bfseries 230} (2021) 9}
  [\href{https://arxiv.org/abs/2012.07971}{{\ttfamily 2012.07971}}].

\bibitem{ms2}
N.E.~Mavromatos and J.~Sol\`a~Peracaula, \emph{{Inflationary physics and
  trans-Planckian conjecture in the stringy running vacuum model: from the
  phantom vacuum to the true vacuum}},
  \href{https://doi.org/10.1140/epjp/s13360-021-02149-6}{\emph{Eur. Phys. J.
  Plus} {\bfseries 136} (2021) 1152}
  [\href{https://arxiv.org/abs/2105.02659}{{\ttfamily 2105.02659}}].

\bibitem{rvm1}
I.L.~Shapiro and J.~Sola, \emph{{On the scaling behavior of the cosmological
  constant and the possible existence of new forces and new light degrees of
  freedom}}, \href{https://doi.org/10.1016/S0370-2693(00)00090-3}{\emph{Phys.
  Lett. B} {\bfseries 475} (2000) 236}
  [\href{https://arxiv.org/abs/hep-ph/9910462}{{\ttfamily hep-ph/9910462}}].

\bibitem{rvm2}
I.L.~Shapiro and J.~Sola, \emph{{Scaling behavior of the cosmological constant:
  Interface between quantum field theory and cosmology}},
  \href{https://doi.org/10.1088/1126-6708/2002/02/006}{\emph{JHEP} {\bfseries
  02} (2002) 006} [\href{https://arxiv.org/abs/hep-th/0012227}{{\ttfamily
  hep-th/0012227}}].

\bibitem{rvmfoss}
J.~Sola, \emph{{Dark energy: A Quantum fossil from the inflationary
  Universe?}}, \href{https://doi.org/10.1088/1751-8113/41/16/164066}{\emph{J.
  Phys. A} {\bfseries 41} (2008) 164066}
  [\href{https://arxiv.org/abs/0710.4151}{{\ttfamily 0710.4151}}].

\bibitem{solaqft}
C.~Moreno-Pulido and J.~Sola, \emph{{Running vacuum in quantum field theory in
  curved spacetime: renormalizing $\rho_{vac}$ without $\sim m^4$ terms}},
  \href{https://doi.org/10.1140/epjc/s10052-020-8238-6}{\emph{Eur. Phys. J. C}
  {\bfseries 80} (2020) 692}
  [\href{https://arxiv.org/abs/2005.03164}{{\ttfamily 2005.03164}}].

\bibitem{rvmevol1}
J.A.S.~Lima, S.~Basilakos and J.~Sola, \emph{{Expansion History with Decaying
  Vacuum: A Complete Cosmological Scenario}},
  \href{https://doi.org/10.1093/mnras/stt220}{\emph{Mon. Not. Roy. Astron.
  Soc.} {\bfseries 431} (2013) 923}
  [\href{https://arxiv.org/abs/1209.2802}{{\ttfamily 1209.2802}}].

\bibitem{rvmevol2}
E.L.D.~Perico, J.A.S.~Lima, S.~Basilakos and J.~Sola, \emph{{Complete Cosmic
  History with a dynamical $\Lambda=\Lambda(H)$ term}},
  \href{https://doi.org/10.1103/PhysRevD.88.063531}{\emph{Phys. Rev. D}
  {\bfseries 88} (2013) 063531}
  [\href{https://arxiv.org/abs/1306.0591}{{\ttfamily 1306.0591}}].

\bibitem{svrcek}
P.~Svrcek and E.~Witten, \emph{{Axions In String Theory}},
  \href{https://doi.org/10.1088/1126-6708/2006/06/051}{\emph{JHEP} {\bfseries
  06} (2006) 051} [\href{https://arxiv.org/abs/hep-th/0605206}{{\ttfamily
  hep-th/0605206}}].

\bibitem{decesare}
M.~de~Cesare, N.E.~Mavromatos and S.~Sarkar, \emph{{On the possibility of
  tree-level leptogenesis from Kalb\textendash{}Ramond torsion background}},
  \href{https://doi.org/10.1140/epjc/s10052-015-3731-z}{\emph{Eur. Phys. J. C}
  {\bfseries 75} (2015) 514} [\href{https://arxiv.org/abs/1412.7077}{{\ttfamily
  1412.7077}}].

\bibitem{boss}
T.~Bossingham, N.E.~Mavromatos and S.~Sarkar, \emph{{The role of temperature
  dependent string-inspired CPT violating backgrounds in leptogenesis and the
  chiral magnetic effect}},
  \href{https://doi.org/10.1140/epjc/s10052-019-6564-3}{\emph{Eur. Phys. J. C}
  {\bfseries 79} (2019) 50} [\href{https://arxiv.org/abs/1810.13384}{{\ttfamily
  1810.13384}}].

\bibitem{mssph}
N.E.~Mavromatos and S.~Sarkar, \emph{{Curvature and thermal corrections in
  tree-level CPT-Violating Leptogenesis}},
  \href{https://doi.org/10.1140/epjc/s10052-020-8109-1}{\emph{Eur. Phys. J. C}
  {\bfseries 80} (2020) 558}
  [\href{https://arxiv.org/abs/2004.10628}{{\ttfamily 2004.10628}}].

\bibitem{sloan}
D.J.~Gross and J.H.~Sloan, \emph{{The Quartic Effective Action for the
  Heterotic String}},
  \href{https://doi.org/10.1016/0550-3213(87)90465-2}{\emph{Nucl. Phys. B}
  {\bfseries 291} (1987) 41}.

\bibitem{metsaev}
R.R.~Metsaev and A.A.~Tseytlin, \emph{{Order alpha-prime (Two Loop) Equivalence
  of the String Equations of Motion and the Sigma Model Weyl Invariance
  Conditions: Dependence on the Dilaton and the Antisymmetric Tensor}},
  \href{https://doi.org/10.1016/0550-3213(87)90077-0}{\emph{Nucl. Phys. B}
  {\bfseries 293} (1987) 385}.

\bibitem{kaloper}
M.J.~Duncan, N.~Kaloper and K.A.~Olive, \emph{{Axion hair and dynamical torsion
  from anomalies}},
  \href{https://doi.org/10.1016/0550-3213(92)90052-D}{\emph{Nucl. Phys. B}
  {\bfseries 387} (1992) 215}.

\bibitem{rvmpheno1}
A.~G\'omez-Valent, J.~Sol\`a and S.~Basilakos, \emph{{Dynamical vacuum energy
  in the expanding Universe confronted with observations: a dedicated study}},
  \href{https://doi.org/10.1088/1475-7516/2015/01/004}{\emph{JCAP} {\bfseries
  01} (2015) 004} [\href{https://arxiv.org/abs/1409.7048}{{\ttfamily
  1409.7048}}].

\bibitem{rvmpheno2}
J.~Sol\`a, A.~G\'omez-Valent and J.~de~Cruz~P\'erez, \emph{{First evidence of
  running cosmic vacuum: challenging the concordance model}},
  \href{https://doi.org/10.3847/1538-4357/836/1/43}{\emph{Astrophys. J.}
  {\bfseries 836} (2017) 43}
  [\href{https://arxiv.org/abs/1602.02103}{{\ttfamily 1602.02103}}].

\bibitem{rvmphenot}
J.~Sol\`a~Peracaula, A.~G\'omez-Valent, J.~de~Cruz~Perez and C.~Moreno-Pulido,
  \emph{{Running vacuum against the $H_0$ and $\sigma_8$ tensions}},
  \href{https://doi.org/10.1209/0295-5075/134/19001}{\emph{EPL} {\bfseries 134}
  (2021) 19001} [\href{https://arxiv.org/abs/2102.12758}{{\ttfamily
  2102.12758}}].

\bibitem{ellis}
J.~Ellis and N.E.~Mavromatos, \emph{{Inflation induced by gravitino
  condensation in supergravity}},
  \href{https://doi.org/10.1103/PhysRevD.88.085029}{\emph{Phys. Rev. D}
  {\bfseries 88} (2013) 085029}
  [\href{https://arxiv.org/abs/1308.1906}{{\ttfamily 1308.1906}}].

\bibitem{lalak}
D.~Coulson, Z.~Lalak and B.A.~Ovrut, \emph{{Biased domain walls}},
  \href{https://doi.org/10.1103/PhysRevD.53.4237}{\emph{Phys. Rev. D}
  {\bfseries 53} (1996) 4237}.

\bibitem{ross}
Z.~Lalak, S.~Lola, B.A.~Ovrut and G.G.~Ross, \emph{{Large scale structure from
  biased nonequilibrium phase transitions: Percolation theory picture}},
  \href{https://doi.org/10.1016/0550-3213(94)00557-U}{\emph{Nucl. Phys. B}
  {\bfseries 434} (1995) 675}
  [\href{https://arxiv.org/abs/hep-ph/9404218}{{\ttfamily hep-ph/9404218}}].

\bibitem{string1}
M.B.~Green, J.H.~Schwarz and E.~Witten, \emph{{Superstring Theory Vol. 1}:
  {25th Anniversary Edition}}, Cambridge Monographs on Mathematical Physics,
  Cambridge University Press (11, 2012),
  \href{https://doi.org/10.1017/CBO9781139248563}{10.1017/CBO9781139248563}.

\bibitem{jackiw}
R.~Jackiw and S.Y.~Pi, \emph{{Chern-Simons modification of general
  relativity}}, \href{https://doi.org/10.1103/PhysRevD.68.104012}{\emph{Phys.
  Rev. D} {\bfseries 68} (2003) 104012}
  [\href{https://arxiv.org/abs/gr-qc/0308071}{{\ttfamily gr-qc/0308071}}].

\bibitem{kerrBH}
B.A.~Campbell, M.J.~Duncan, N.~Kaloper and K.A.~Olive, \emph{{Axion hair for
  Kerr black holes}},
  \href{https://doi.org/10.1016/0370-2693(90)90227-W}{\emph{Phys. Lett. B}
  {\bfseries 251} (1990) 34}.

\bibitem{stephon}
S.H.-S.~Alexander, M.E.~Peskin and M.M.~Sheikh-Jabbari, \emph{{Leptogenesis
  from gravity waves in models of inflation}},
  \href{https://doi.org/10.1103/PhysRevLett.96.081301}{\emph{Phys. Rev. Lett.}
  {\bfseries 96} (2006) 081301}
  [\href{https://arxiv.org/abs/hep-th/0403069}{{\ttfamily hep-th/0403069}}].

\end{thebibliography}\endgroup

\end{document}